\documentclass[runningheads]{llncs}
\usepackage[T1]{fontenc}
\usepackage{graphicx}
\usepackage{xurl}
\usepackage{hyperref}

\usepackage{color}

\begin{document}

\title{Narrative Memory in Machines: Multi-Agent Arc Extraction in Serialized TV}
\titlerunning{Multi-Agent Arc Extraction in Serialized TV}
%
\author{Roberto Balestri\inst{1}\orcidID{0009-0000-5008-2911}\\ \and
Guglielmo Pescatore\inst{2}\orcidID{0000-0001-5206-6464}
}
\authorrunning{R. Balestri and G. Pescatore}
%
\institute{Department of the Arts, Università di Bologna, Italy\\
\email{roberto.balestri2@unibo.it}\\
\and
Department of the Arts, Università di Bologna, Italy\\
\email{guglielmo.pescatore@unibo.it}
}

\maketitle              
\textit{Preprint of the forthcoming paper to appear in Lecture Notes in Artificial Intelligence, scheduled for publication on December 5, 2025.}

\abstract{Serialized television narratives present significant analytical challenges due to their complex, temporally distributed storylines that necessitate sophisticated information management. This paper introduces a multi-agent system (MAS) designed to extract and analyze narrative arcs by implementing principles of computational memory architectures. The system conceptualizes narrative understanding through analogues of human memory: Large Language Models (LLMs) provide a form of semantic memory for general narrative patterns, while a vector database stores specific arc progressions as episodic memories. A multi-agent workflow simulates working memory processes to integrate these information types. Tested on the first season of \textit{Grey's Anatomy} (ABC 2005-), the MAS identifies three arc types: Anthology (self-contained), Soap (relationship-focused), and Genre-Specific. These arcs and their episodic developments are stored in a vector database, facilitating structured analysis and semantic comparison. To bridge automation with critical interpretation, a graphical interface enables human oversight and refinement of the system's narrative memory. While demonstrating strong performance in identifying Anthology Arcs and character entities, the system's reliance on textual paratexts (episode summaries) revealed limitations in discerning overlapping arcs and opaque dynamics, underscoring the challenges in computational memory consolidation versus human holistic understanding. This memory-centric approach highlights the potential of combining AI-driven memory processing with human expertise. Beyond television, it offers promise for serialized written formats where narrative is entirely text-based. Future work will focus on integrating multimodal inputs to enrich episodic memory, refining memory integration mechanisms within the MAS, and expanding testing across diverse genres.}


\section{\uppercase{Introduction}}
\label{sec:introduction}

The analysis of narrative structure, a long-standing pursuit in literary and media studies, confronts unique hurdles with the advent of serialized television. These series weave intricate storylines across numerous episodes and seasons, demanding sophisticated methods for tracking and comprehension \cite{mittell2015complex}. Understanding such narratives can be framed as a complex memory management problem, where viewers, much like computational systems, must store, retrieve, and integrate vast amounts of information over extended temporal periods. The challenge intensifies as narrative arcs intertwine, develop non-linearly, and require audiences to maintain an evolving mental model of the story world. This parallels the cognitive load experienced by humans when processing and recalling lengthy, complex event sequences \cite{zacks2007event}.

This paper builds upon our previous work, in which we introduced a multi-agent system (MAS) for narrative arc extraction \cite{balestri2025multi}. In this new contribution, we reconceptualize narrative arc extraction through the framework of computational memory architectures, drawing analogies with human cognitive memory functions. We posit that effective narrative analysis, particularly for serialized content, can benefit from principles underlying declarative memory, which in humans encompasses both episodic memory (specific, contextualized experiences) and semantic memory (general knowledge) \cite{tulving1998episodic}.

Our system reflects this dual structure: a Large Language Model (LLM) serves as a semantic memory component, drawing on general narrative conventions and world knowledge, while specific narrative events—parsed from episode summaries and organized into structured progressions—are stored in a vector database. Crucially, while these progressions provide episode-level narrative context, the atomic units of episodic memory in our model are the utterances within each progression, each representing a discrete event or character interaction. This granularity allows the system to retrieve and reason over narrative content with a level of precision analogous to human recall of detailed, context-rich occurrences.

The sequential processing and collaborative decision-making of the specialized agents within the MAS simulate a computational working memory. Each agent actively manipulates and integrates information retrieved from both episodic and semantic stores, constructing and updating a coherent model of the evolving narrative.

Serialized narratives, such as the first season of \textit{Grey's Anatomy} (ABC 2005-) used in our testing, present a significant challenge due to their "distributed memory" demands – storylines unfold across multiple episodes, requiring viewers (and analytical systems) to recall and connect disparate pieces of information \cite{mittell2010previously}. Our multi-agent system is designed to address this by systematically identifying, categorizing, and mapping these narrative arcs. It parses episode summaries to extract arcs, storing their episodic developments in both relational and vector databases, thereby creating a structured and semantically rich representation of the narrative's memory. This approach aims to bridge the gap between the automated processing of narrative data and the nuanced interpretation required for comprehensive narratological analysis, offering a novel framework for computational memory in the study of serialized storytelling.

The remainder of this paper is organized as follows.
Section~\ref{sec:related_works} reviews related work in narrative analysis, computational methods, memory systems, and serialized media.
Section~\ref{sec:theoretical_framework} details the theoretical framework of computational memory architectures (episodic, semantic, working) for narrative analysis.
Section~\ref{sec:narrative_arcs} defines the model for narrative arc types and their episodic progressions for memory storage.
Section~\ref{sec:technical} outlines the technical infrastructure, including the LLM, databases, embedding models, and other software tools used.
Section~\ref{sec:material_collection} describes material collection and preprocessing, including cleaning, simplification, and entity normalization.
Section~\ref{sec:mas_extraction} presents the multi-agent system for arc extraction, its memory-informed architecture, agent workflow, and semantic comparison processes.
Section~\ref{sec:graphic} details the graphical interface for human oversight, enabling visualization, editing, and refinement of the narrative memory.
Section~\ref{sec:test_improv} evaluates system performance against human analysis, assessing character/arc extraction and memory system functioning.
Section~\ref{sec:discussion_memory} discusses theoretical contributions, computational versus human narrative memory, and implications for computational narratology.
Finally, Section~\ref{sec:conclusion} summarizes findings, limitations, and future work, including multimodal inputs and enhanced memory mechanisms.

The repository containing the code is available on Github at the following link \url{https://github.com/robertobalestri/MAS-AI-Assisted-Narrative-Arcs-Extraction-TV-Series}.

\section{\uppercase{Related Works}}
\label{sec:related_works}

The intersection of narrative analysis, memory systems, and computational methods represents a convergent evolution across multiple domains. From the structural analysis of folktales to contemporary memory-augmented artificial intelligence, researchers have increasingly recognized that effective information management—whether in human cognition, serialized storytelling, or computational systems—relies on similar architectural principles for storing, retrieving, and activating complex information.

\subsection{Memory Architectures and Information Processing}

The foundational work of \cite{tulving1998episodic} established the crucial distinction between episodic memory, which stores specific contextualized experiences, and semantic memory, which maintains generalized knowledge structures. This dichotomy has proven remarkably persistent across domains, manifesting in how television narratives distinguish between particular story events and general world-building frameworks \cite{innocenti2017narrative}, and how Retrieval-Augmented Generation (RAG) systems separate specific document retrieval from parametric knowledge representation.

\cite{baddeley2003working} expanded this framework by identifying working memory as the critical bottleneck where archived information becomes available for active processing. This limitation—managing vast stored knowledge through constrained processing capacity—emerges as a universal constraint across biological, narrative, and computational systems. Human cognition faces this challenge through selective attention and associative recall, television series through strategic redundancy and narrative cues \cite{mittell2010previously}, and AI systems through context window management and retrieval optimization.

Cognitive science has revealed how declarative memory systems enable long-term information coherence while maintaining adaptive flexibility \cite{squire1996structure}. These insights prove particularly relevant for understanding how serialized narratives manage extended storytelling and how memory-augmented AI systems maintain contextual awareness across interactions.

\subsection{Computational Approaches to Narrative Understanding}

Early computational narrative analysis relied on traditional machine learning approaches, including Support Vector Machines \cite{hearst1998support} and logistic regression \cite{vimal2020application}, which struggled to capture the temporal dynamics and complex relationships inherent in extended narratives \cite{young2018recent}. The emergence of deep learning architectures, particularly recurrent neural networks and convolutional networks, marked significant progress in modeling sequential patterns and local narrative features \cite{yin2017comparative}.

The transformer revolution, exemplified by models such as GPT \cite{radford2018improving} and BERT \cite{vaswani2017attention}, introduced self-attention mechanisms capable of capturing long-range dependencies in textual narratives. These developments enabled more sophisticated semantic embeddings \cite{haywood2024tokens} that represent narrative relationships mathematically, facilitating deeper understanding of story structures and character dynamics \cite{esposti2023exploring}.

RAG systems represent a particularly significant evolution, combining parametric knowledge with external document retrieval \cite{lewis2020retrieval}. This architecture mirrors the episodic-semantic memory distinction: language models provide general linguistic and world knowledge (semantic memory), while vector databases store specific, contextualized documents that can be retrieved and activated for particular tasks (episodic memory).

Contemporary work in memory-augmented systems has explored various approaches to managing external knowledge, from simple retrieval mechanisms to sophisticated attention-based integration \cite{wang2025m+}. These developments align with our understanding of how human memory systems balance storage efficiency with retrieval effectiveness \cite{baddeley2003working}.

\subsection{Narrative Structure and Temporal Dynamics in Serialized Media}

The study of narrative structures has evolved from classical formalist approaches \cite{propp2012russian,todorov1969structural,bann1973russian} toward understanding complex, temporally distributed storytelling systems. Television series pose unique challenges that distinguish them from traditional narrative forms, requiring analytical frameworks capable of tracking interconnected storylines across extended temporal spans.

\cite{mittell2010previously} identified how television series function as "memory systems," requiring viewers to manage accumulated narrative knowledge while processing new information. This work introduced crucial concepts including "formal memory"—viewers' knowledge of series-specific narrative conventions—and "surprise memory," where narrative pleasure derives from the cognitive process of remembering rather than learning new information. These insights reveal how serialized media must actively manage audience memory states, balancing accessibility for newcomers with sophistication for experienced viewers.

Recent analysis of emotional arcs and thematic progressions \cite{reagan2016emotional,schmidt2015plot} has demonstrated how computational methods can identify patterns in narrative development, though capturing non-linear temporal dynamics and complex turning points remains challenging \cite{piper2023quantitative}. Concepts such as narrative flow, suspense, and causality require sophisticated approaches that can model both local events and global narrative coherence \cite{wilmot-keller-2020-modelling,graesser1994constructing}.

The multi-layered structure of contemporary television, operating through networks of interconnected storylines rather than single linear plots \cite{perez2021multi}, necessitates analytical frameworks capable of tracking multiple concurrent narrative threads. This complexity mirrors challenges in memory management systems, where multiple information streams must be maintained and selectively activated based on contextual demands.

\subsection{Multi-Agent Systems and Collaborative Intelligence}

Multi-Agent Systems (MAS) have emerged as powerful frameworks for complex task decomposition, enabling specialized agents to collaborate on problems that exceed individual processing capabilities. Applications in legal reasoning \cite{yuan2024can} and scientific theory development \cite{ghafarollahi2024sciagents} demonstrate how distributed processing can enhance analytical depth and reliability.

In narrative analysis, MAS approaches have shown particular promise for managing the multifaceted nature of storytelling \cite{aoki2023analysis,balestri2025multi}. The ability to assign specialized functions to individual agents—character extraction, temporal analysis, thematic identification—while maintaining system-wide coherence aligns well with how human analysts naturally decompose complex narrative understanding tasks.

The integration of MAS with memory-augmented architectures represents a convergent evolution toward systems that can maintain both specialized expertise and integrated understanding. This approach mirrors how human cognitive systems employ distributed processing while maintaining a unified conscious experience \cite{minsky1986society}, suggesting that effective artificial systems may need to implement similar architectural principles.

\subsection{Character Networks and Relationship Dynamics}

Recent work in computational narrative analysis has focused on extracting and modeling character relationships and interactions \cite{dalla20233,janosov2021network,bost2016narrative,beveridge2018game,innocenti2018evolution}.

The challenge of character entity recognition in serialized media extends beyond simple name extraction to include handling alternative appellations, temporal character development, and relationship dynamics that evolve across episodes. This problem parallels challenges in memory systems, where the same conceptual entity may be represented differently across contexts while maintaining underlying identity \cite{schank1982theory}.

Network analysis approaches have revealed how character relationships form the structural backbone of serialized narratives, with relationship dynamics often serving as the primary drivers of long-term story arcs \cite{liu2017balance,min2019modeling}. Understanding these patterns requires systems capable of tracking both explicit interactions and implicit relationship changes over extended temporal spans.

\section{\uppercase{Theoretical Framework: Memory Architectures in Computational Narrative Analysis}}
\label{sec:theoretical_framework}

Analyzing serialized narratives requires sophisticated memory management. This section frames our multi-agent system's design for narrative arc extraction by implementing computational memory architectures—episodic, semantic, and working memory—analogous to human cognitive systems.

\subsection{Declarative Memory in Narrative Systems}
Our system mirrors human declarative memory principles \cite{tulving1998episodic} for managing narrative information:

\begin{itemize}
\item \textbf{Episodic Memory in Narrative Computation}: Corresponds to human memory for specific, contextualized events \cite{tulving1998episodic}. Our system stores narrative arc progressions from specific episodes as episodic traces. A vector database, holding embeddings of these progressions, facilitates semantic retrieval of these narrative "episodes."

\item \textbf{Semantic Memory in Narrative Computation}: Analogous to human general knowledge \cite{tulving1998episodic}, the LLM acts as the semantic memory. Its pre-trained understanding of narrative conventions, archetypes, and linguistic patterns is vital for interpreting plots and identifying arcs.

\item \textbf{Working Memory in Narrative Computation}: Parallels human temporary information processing \cite{baddeley2003working}. The multi-agent system's sequential pipeline functions as a computational working memory, where agents actively process current episode data by integrating information from the episodic (vector database) and semantic (LLM) stores.

\end{itemize}

\subsection{Information Storage and Retrieval Patterns}
The system's memory efficacy hinges on robust storage and efficient retrieval, primarily centered on the vector database. This \textbf{semantic storage} holds embeddings of arc titles, descriptions, and progression content. Retrieval mechanisms then utilize semantic similarity searches within this vector database to identify related past arcs. This process models associative recall, crucial for discerning evolving storylines even when phrasings differ \cite{lewis2020retrieval}. A relational database provides ancillary support, mainly for managing structured data such as character names and essential metadata, ensuring organized factual reference. Activation patterns for retrieved information are managed by the MAS, which determines the relevance of stored data for the current episode's analysis.

\subsection{Memory-Guided Multi-Agent Architecture}
The multi-agent architecture is not merely a task-decomposition strategy but a framework for distributed memory processing:
\begin{itemize}
    \item Each agent can be seen as a specialized memory processor, focusing on specific aspects of narrative information (e.g., identifying new episodic events for Anthology Arcs, or tracking long-term relationship developments for Soap Arcs; see \ref{sec:narrative_arcs} for additional details).
    \item The sequential workflow, where agents build upon the output of previous agents and incorporate information from prior episodes, simulates a process of memory consolidation and updating. Agent 1 (Existing Season Arcs Identifier), for instance, explicitly queries the system's "memory" of arcs from previous episodes.
    \item The integration of stored narrative knowledge (episodic traces from the vector DB) with the LLM's general narrative understanding (semantic knowledge) within each agent's decision-making process is key to the system's ability to interpret new information in the context of past events.
\end{itemize}
This memory-centric view of the MAS highlights how it attempts to computationally model the dynamic processes of recall, recognition, and integration that are fundamental to comprehending serialized narratives.

\section{\uppercase{Modeling Narrative Arcs for Storing}}
\label{sec:narrative_arcs}

To analyze and document these arcs effectively, we conceptualized a model that captures their key attributes and episodic progressions in a structured format, which is intrinsically linked to how narrative information is committed to and retrieved from the system's memory stores. Each arc, taking inspiration from \cite{pescatore2019narration,rocchi2022}, is categorized into one of three types based on its nature and scope:
\begin{itemize}
    \item Anthology Arcs, which are self-contained stories resolved within a single episode, often centered on genre-specific events or cases. These represent distinct, easily delineable episodic memories.
    \item Genre-Specific Arcs, focusing on professional or thematic elements tied to the series, such as workplace dynamics or medical conflicts. These may span multiple episodes, requiring the linking of related episodic memories.
    \item Soap Arcs, which explore interpersonal relationships, personal growth, and emotional conflicts. These may span multiple episodes, evolving gradually over time.
\end{itemize}

\begin{figure}
\centering
\includegraphics[width=\textwidth]{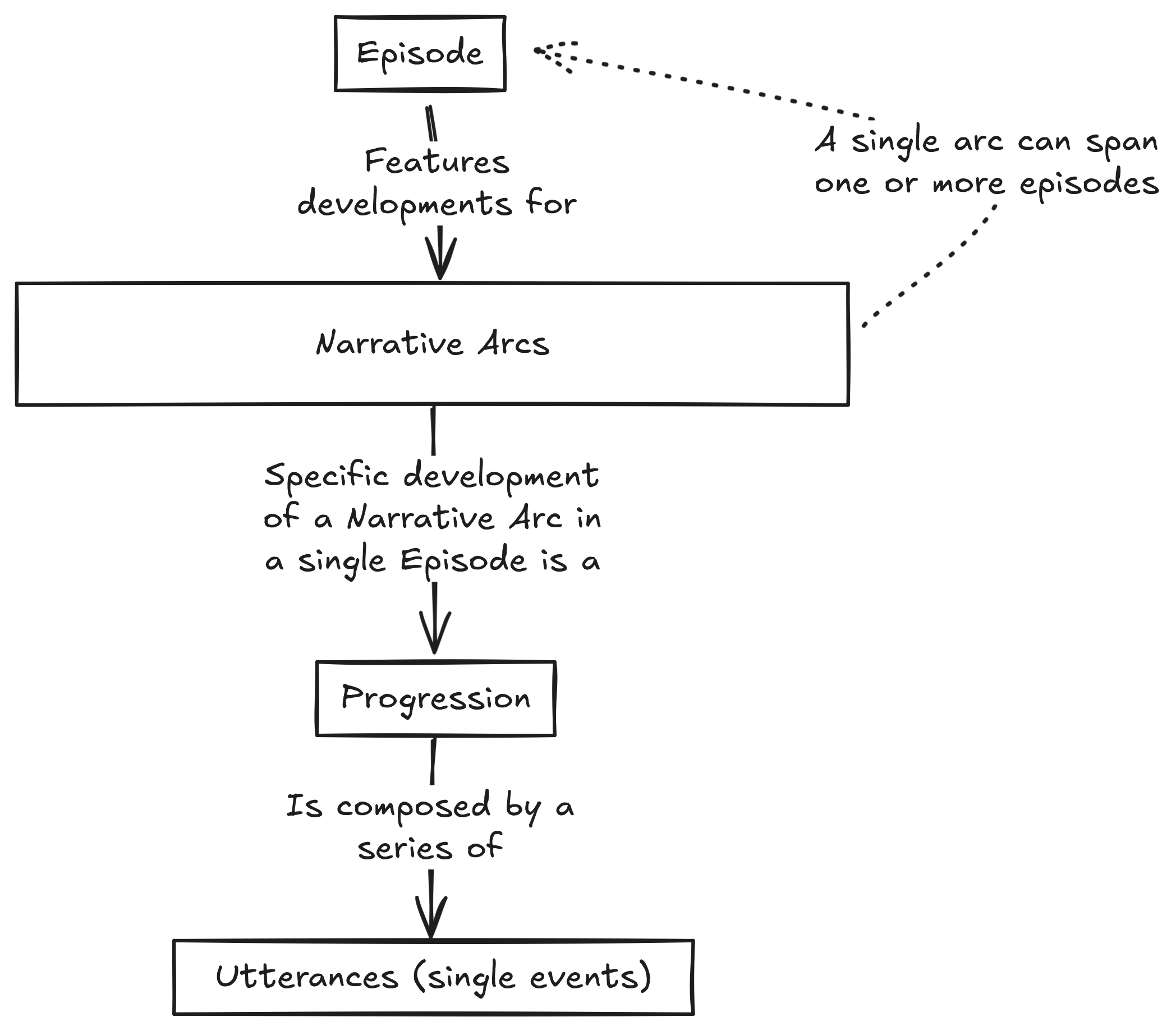}
\caption{Diagram illustrating the hierarchical structure of Arcs, Progressions, and Utterances.}
\label{fig:hierarc}
\end{figure}

Each arc is defined by a title and description that encapsulates its central theme or conflict. The main characters driving the arc are identified, along with the arc type, which provides context for its role within the series.
To track how arcs evolve over time, we document their progressions, which capture developments relevant to each storyline along one series' episode. A progression specifies the episode and season where it occurs, the interfering characters influencing the arc in that specific context, and a concise description of the events advancing the storyline. Each of these progressions may consist of chronologically ordered utterances, each operating as a distinct unit of episodic memory within the broader narrative arc \cite{tulving1998episodic}. See Figure \ref{fig:hierarc} for a straightforward view of the hierarchy.

Summarizing, in our system, the object \textit{Narrative Arc} includes several fields. Each arc is assigned a unique identifier (\textit{arc\_id}), along with a title and description encapsulating its central theme or conflict. The arcs also contain progressions, which are a list of \textit{Progression} objects representing their developments across episodes. Additionally, each arc identifies the main characters driving the storyline, specifies the arc type (Anthology, Soap, or Genre-Specific), and indicates the series to which it belongs. The database ensures this structured information is maintained consistently, akin to a well-organized long-term memory store.

Each \textit{Progression} is a structured object that captures a set of narrative developments occurring within a specific episode of a television series. It is assigned a unique identifier (\textit{progression\_id}) and is linked to its parent narrative arc via the \textit{arc\_id}. Alongside metadata indicating the series, season, and episode, the progression contains a field called \textit{Content}, which includes a short, chronologically ordered sequence of phrases (utterances) summarizing key developments relevant to that arc in the given episode.

These progressions are embedded and indexed within a vector database, allowing for semantic retrieval and comparison. However, it is crucial to emphasize that in our system, the \textit{Progression} as a whole does not constitute the elementary unit of episodic memory. Rather, each utterance within a progression—typically corresponding to a distinct event, action, or turning point in the narrative—is treated as a discrete memory trace. This granularity is consistent with cognitive models of episodic memory, which emphasize the storage of fine-grained, event-level information rather than coarse-grained aggregates. Consequently, a single progression functions as a composite memory structure, grouping together multiple event-level memory units under a shared temporal and narrative frame.

\section{\uppercase{Technical Details}}
\label{sec:technical}
In this paper, "LLM" refers specifically to the OpenAI GPT-4o model, accessed via API. This model serves as the primary source of semantic memory, providing general world knowledge and narrative understanding \cite{radford2018improving}. Data storage was implemented using an SQLite database for structured relational data, forming one part of the system's long-term memory. Embeddings, crucial for the system's episodic memory component, were generated with the Cohere-embed-english-v3.0 model and stored in a Chroma vector database. The backend was developed in Python (version 3.11), with FastAPI facilitating communication between the backend and frontend, which was built using the React JavaScript framework. Character entities were extracted from episode plots using the Spacy NLP model \textit{en\_core\_web\_trf} and refined by a custom LLM-driven script.

\section{\uppercase{Material Collection and Preprocessing}}
\label{sec:material_collection}

Our software was designed to be generalizable across various TV series genres. For testing, we focused on the first season of \textit{Grey's Anatomy}, a choice motivated by our team’s prior research on medical dramas. This familiarity provided a solid foundation for evaluating the software’s performance. The system requires only episode summaries of a season to function effectively. This stage can be viewed as "memory preparation," ensuring that the information encoded into the system's memory is clear, structured, and optimized for later retrieval and processing.

\subsection{Episodes' Plots Gathering}
To develop and test the software, we sourced episode summaries from the fan-maintained \textit{Grey's Anatomy Wiki} (\url{https://greysanatomy.fandom.com}), which operates under the Creative Commons Attribution-Share Alike License (CC BY-SA). This license permits research and commercial use with proper attribution. The raw data was preprocessed to ensure standardization and usability for analysis.

\subsection{Data Cleaning and Simplification}
The preprocessing began with cleaning and simplifying the episode summaries. Using the LLM (our semantic memory resource), sentences were rewritten in simpler, more structured forms, emphasizing clarity by focusing on a single character or event per sentence. Direct quotations were avoided, and complex sentence structures were replaced with concise descriptions. This process ensured the plots were easily interpretable and consistent, improving the quality of the episodic information to be stored.

\subsection{Character Entity Normalization}
Character name variability was addressed by replacing pronouns with corresponding character names, guided by contextual analysis within a fifteen-sentence window using the LLM. The spaCy Transformer model was used to extract character entities, and the LLM further refined this output by resolving similar names and alternative appellations. A character database was created, containing unique identifiers, preferred names, and alternative names for each character. The plots were then standardized by replacing all character references with their preferred names. This normalization is crucial for maintaining coherent "memory traces" for each character across episodes.

The quality of this "encoded" information directly impacts the performance of the memory retrieval and integration processes later on.

\section{\uppercase{The Multi-Agent System for Arc Extraction}}
\label{sec:mas_extraction}

\begin{figure}
\centering
\includegraphics[width=\textwidth]{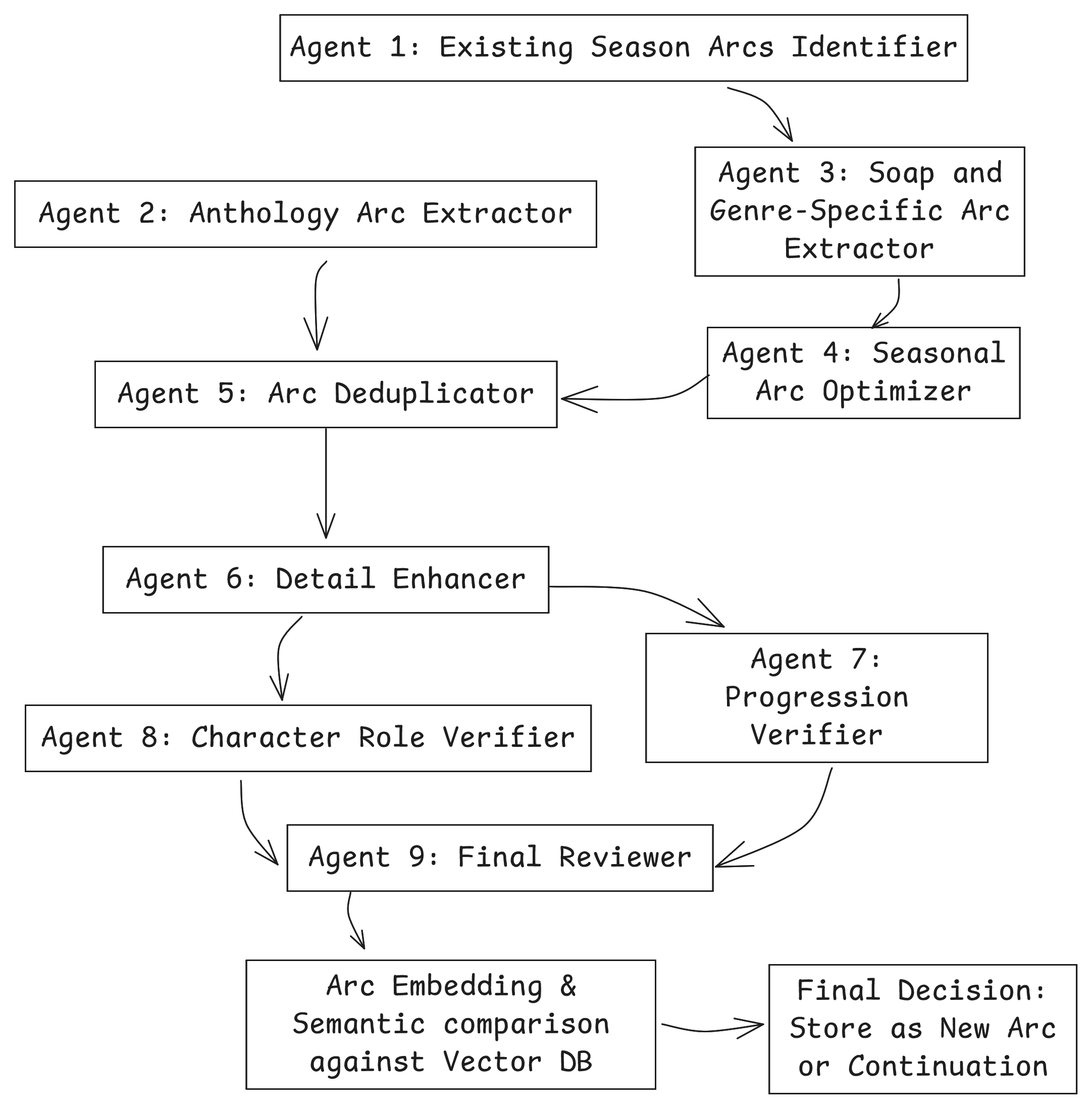}
\caption{Narrative Arc Extraction Process as seen in \cite{balestri2025multi}}
\label{fig:extraction}
\end{figure}

The narrative arc extraction process utilizes a multi-agent system to analyze the episode plots. This system is designed not just for task decomposition but as a dynamic memory processing architecture. It identifies, categorizes, and refines narrative arcs, integrating episodic storylines into a cohesive seasonal framework by leveraging both episodic (vector database) and semantic (LLM) memory components. The workflow is sequential, with specialized agents performing tasks that often involve querying, updating, or comparing information within these memory stores. The resulting arcs and their progressions are stored in a vector database, as detailed in Section \ref{sec:semantic_comparison}. Figure \ref{fig:extraction} provides a visual representation of the system.

\subsection{Memory-Informed System Architecture}
The MAS architecture is inherently memory-informed. Each agent's operation relies on accessing and manipulating different facets of the system's memory:
\begin{itemize}
    \item \textbf{Episodic Memory Access}: Agents frequently query the vector database to retrieve semantically similar past arcs or progressions. This is crucial for Agent 1 (Existing Season Arcs Identifier) and Agent 4 (Seasonal Arc Optimizer) to determine if current events are continuations of, or related to, previously stored narrative episodes.
    \item \textbf{Semantic Memory Utilization}: The LLM is consulted by multiple agents (e.g., Agent 2 for Anthology Arcs, Agent 3 for Soap/Genre-Specific Arcs, Agent 6 for Detail Enhancement) to interpret plot summaries, generate descriptions, categorize arcs, and infer relationships based on its general narrative understanding.
    \item \textbf{Working Memory Simulation}: The sequential flow of information between agents, where the output of one becomes the input for the next, simulates a working memory process \cite{baddeley2003working}. Each agent processes a limited amount of information (the current episode's plot, plus relevant retrieved memories) to make its decisions.
    \item \textbf{Memory Integration}: A key function is the integration of information from these memory stores. For example, Agent 4 must compare newly identified arcs (derived from current episode processing and semantic interpretation by the LLM) with existing arcs retrieved from episodic memory to avoid redundancy and ensure coherence.
\end{itemize}

\subsection{Workflow Design}
The extraction workflow is divided into stages, with each stage handled by an autonomous agent. Each agent focuses on a specific aspect of narrative analysis, ensuring arcs are both episode-specific and seasonally coherent. Results from prior episodes, retrieved from the system's episodic memory, are incorporated to maintain continuity and account for the evolving nature of serialized storytelling.

\subsubsection{Agent 1 - Existing Season Arcs Identifier}
This agent queries the system's episodic memory (vector database) to evaluate if arcs from previous episodes are present in the current episode. If semantically similar arcs are detected, they are flagged as "possibly present," activating these memory traces for consideration by subsequent agents.

\subsubsection{Agent 2 - Anthology Arc Extractor}
Leveraging the LLM's semantic understanding of narrative structure, this agent identifies self-contained, standalone storylines unique to the current episode. These are treated as distinct episodic memories, as they typically do not require extensive linking to past arc memories.

\subsubsection{Agent 3 - Soap and Genre-Specific Arc Extractor}
This agent analyzes the episode plot, using the LLM's semantic capabilities, to identify new soap and genre-specific arcs. It also validates arcs flagged by Agent 1 by comparing the current episode's events against the retrieved episodic information of those past arcs.

\subsubsection{Agent 4 - Seasonal Arc Optimizer}
This agent performs a critical memory consolidation task. It minimizes redundancy by analyzing soap and genre-specific arcs for overlaps. It performs stricter checks on "possibly present" season arcs (retrieved from episodic memory) and newly identified arcs, merging or refining them as needed. This ensures that distinct narrative memory streams are maintained without unnecessary duplication.

\subsubsection{Agent 5 - Arc Deduplicator}
All arcs extracted from the episode, including Anthology Arcs, are reviewed for similarity. This agent acts as a final check on the working memory's contents for the current episode, resolving overlaps through a disambiguation process, often guided by the LLM's semantic judgment. For example, arcs identified as both Anthology and Genre-Specific by separate agents are clarified.

\subsubsection{Agent 6 - Detail Enhancer}
This agent enriches each arc with detailed contextual information, effectively elaborating the episodic memory trace. This includes: main characters driving the arc, supporting or interfering characters influencing the arc in the episode, a concise description of the arc’s episodic events (the "progression"). This relies on the LLM to synthesize information from the plot.

\subsubsection{Agent 7 - Progression Verifier}
The progressions of each arc (the core of the episodic memory for that arc in that episode) are reviewed to ensure specificity and relevance, again using the LLM's semantic understanding. This step ensures that significant developments are captured without overlapping with unrelated narratives.

\subsubsection{Agent 8 - Character Role Verifier}
This agent verifies and, if necessary, corrects the classification of characters as either main or interfering, ensuring the "cast list" for each episodic memory is accurate.

\subsubsection{Agent 9 - Final Reviewer}
A final verification ensures narrative consistency. Validated arcs are then committed to long-term storage: the relational database for their structured data and the vector database for their semantic embeddings (updating the system's episodic memory store for future retrieval).

\subsection{Semantic Comparison and Arc Embedding}
\label{sec:semantic_comparison}
After analyzing the single episode, embeddings are generated for the arcs and their progressions. This process transforms textual narrative information into a format suitable for the system's episodic memory (the vector database).

These embeddings, for a final deduplication, are compared against the vector database to identify semantically similar arcs. If similarities are detected, an LLM determines whether the arcs represent the same storyline or not. Based on this determination, the system either stores them as new arcs (new, distinct episodic memory chains) or links them as continuations of overarching plotlines (extending existing episodic memory chains).

\section{\uppercase{Graphic Interface}}
\label{sec:graphic}

Human input remains essential for refining the outputs of the multi-agent system, particularly given the nuances of narrative interpretation that can challenge purely computational memory systems \cite{piper2019enumerations}. The graphical interface is designed to facilitate this oversight and correction, acting as a "memory visualization and manipulation tool." It balances simplicity with advanced features, enabling users to visualize, edit, regenerate, and analyze narrative arcs effectively, thereby allowing for human refinement of the computationally generated narrative memory.

\subsection{Interface Overview}

The interface is divided into three main sections:
\begin{itemize}
\item Narrative Arc Timeline: Provides a visual representation of how storylines (mnarrative arcs) evolve across episodes in a season. Users can apply filters based on arc type (Anthology, Genre-Specific, Soap) or associated characters. This allows users to inspect the temporal continuity of the system's narrative memory.
\item Vector Store Explorer: Includes tools for visualizing arc embeddings (the system's episodic memories) using 3D PCA and displays clusters of similar arcs. This offers insight into how the system is associating different narrative events.
\item Character Section: Allows users to explore extracted character entities, manage appellations, or merge similar characters, ensuring consistency in the "actors" within the narrative memories.
\end{itemize}

\subsection{Key Features and Functionalities}

\subsubsection{Arc and Progression Visualization}
\begin{figure}
\centering
\includegraphics[width=\textwidth]{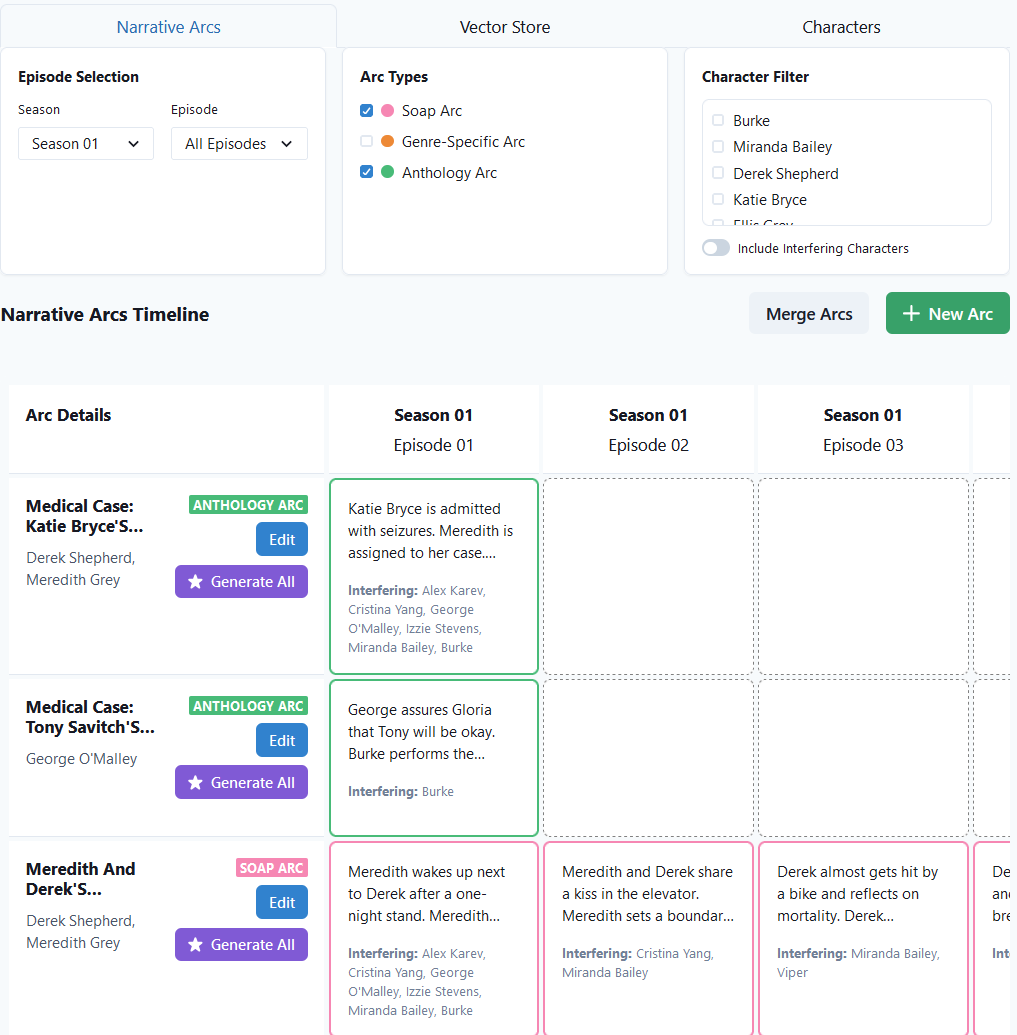}
\caption{The main view of the graphical interface, displaying narrative arcs and their episodic progressions. As seen in \cite{balestri2025multi}.}
\label{fig:main_view}
\end{figure}

Users can navigate narrative arcs through a tabular interface, where rows represent arcs (memory chains), and columns correspond to episodes. Each cell displays the arc’s progression (a specific episodic memory) within a specific episode (see Figure \ref{fig:main_view}).

\subsubsection{Arc Creation and Editing}
Arcs can be created or edited via a dialog box. Users specify attributes such as title, description, and arc type, as well as main characters and episodic progressions. This allows users to manually correct or supplement the system's memory. Progressions belonging to a given arc (relative to one or more episodes) can also be auto-generated using AI for efficiency.

\subsubsection{Arc Merging and Deduplication}
For overlapping or semantically similar arcs identified by the system (or by the user), users can compare them side by side and merge duplicates. This is a critical memory consolidation step, guided by human intelligence, to maintain consistency in the narrative memory.

\subsubsection{Progression Management}
Users can manually create or edit episodic progressions, which include episode-specific descriptions and interfering characters. Alternatively, progressions can be auto-generated and refined as needed. This provides fine-grained control over the content of individual episodic memories.

\subsubsection{Clustering and Semantic Analysis}
\begin{figure}
\centering
\includegraphics[width=\textwidth]{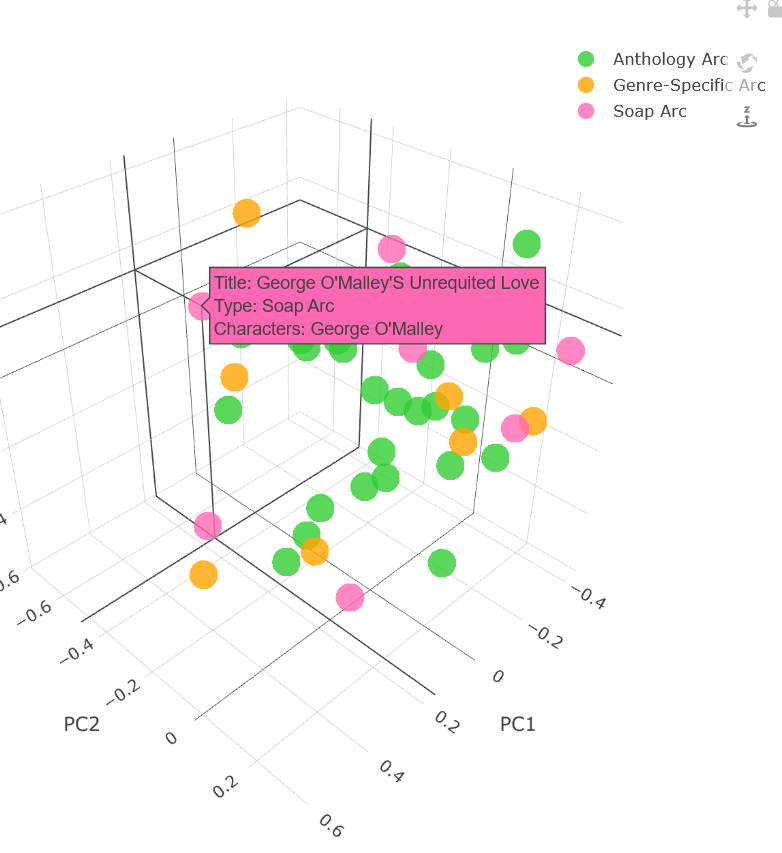}
\caption{3D PCA visualizer for clustering arcs based on semantic similarity, offering a view into the episodic memory's structure. As seen in \cite{balestri2025multi}}
\label{fig:3d_pca}
\end{figure}

Clustering tools group arcs based on semantic embeddings (retrieved from the episodic memory store) and display them in both tabular and 3D PCA formats (see Figure \ref{fig:3d_pca}). This feature helps users identify thematic connections and detect arcs that the multi-agent system incorrectly treated as distinct, essentially allowing users to explore the associative structure of the system's narrative memory.

\subsubsection{Character Management}
The character panel enables users to view, edit, and merge characters to ensure consistency across arcs. A similarity threshold based on the Jaccard index \cite{niwattanakul2013using} highlights potential duplicate characters for resolution.

\section{\uppercase{Experimental Evaluation and Memory System Performance}}
\label{sec:test_improv}

\subsection{Test Setup}

The system’s performance was evaluated by comparing its extracted narrative arcs, based solely on episode summaries (paratexts), with arcs identified by a human scholar. This human analysis, derived from watching each episode of \textit{Grey's Anatomy}'s first season at least twice, serves as a gold standard representing a deep, contextually rich "human memory" of the narrative. The evaluation focuses not just on raw extraction accuracy but on how well the system's memory components (episodic, semantic, working) function in tandem.

Paratexts \cite{genette1997paratexts}, such as episode plots, fan fiction, plot maps, and similar materials, have gained increasing significance in the Internet era \cite{jenkins2012textual,gray2017fandom}. While these materials do not, by their intrinsic nature, constitute the primary text itself, they provide scholars like us with valuable supplementary resources for research and analysis. For our system, these paratexts serve as the primary input for constructing its narrative memory.

The test was conducted across the entire season, and the results were analyzed to assess the system's capabilities in memory storage, integration, and retrieval for narrative analysis.

\subsection{Results and Memory System Performance}

\subsubsection{Pros: Character Entities and Anthology Arcs}
The system demonstrated notable strength in its ability to perform LLM-aided character entity recognition and linking. It successfully identified 62 character entities within the test dataset, with 61 of these being correct. The single duplicate entry ("Frost" and "Jerry Frost") resulted from inconsistent naming in the source material, a challenge for memory systems relying on surface forms. Implementing a more sophisticated memory reconciliation mechanism, perhaps a two-agent verification system for entities, could resolve such errors but would increase computational costs associated with working memory operations.

In arc extraction, the system excelled in identifying Anthology Arcs, achieving a precision of 89.3\% (25 out of 28 arcs correctly extracted). Anthology Arcs, being self-contained within single episodes, represent simpler sets of episodic memories. Their clear boundaries and limited temporal extent make them easier to detect.

\subsubsection{Cons: Coherence Challenges}
Challenges were observed with Soap and Genre-Specific arcs, which require more complex memory integration across multiple episodes and the linking of various episodic memory traces into coherent, long-term narratives. The system occasionally duplicated arcs or failed to merge arcs representing the same storyline. For example, the arcs:
\begin{itemize} \item "Izzie Stevens: Overcoming Past and Professional Growth" \item "Izzie Stevens: Balancing Personal Life and Professional Ambitions" \end{itemize}
were treated as separate memory streams by the system. Human analysis, however, integrated these into a unified storyline, suggesting a more holistic memory consolidation process. This indicates a limitation in the system's current ability to recognize that different facets of a character's development can belong to a single, overarching remembered narrative.

Conversely, the system overlooked the shared arc of Meredith Grey and Derek Shepherd’s relationship. Instead, it identified only individual character arcs for Meredith and Derek. This "dilution" of a joint storyline into separate individual memory traces highlights a difficulty in forming and retrieving memories of shared or relational narratives from the provided paratexts. The episodic information might not have been salient enough in the summaries to trigger the formation of a distinct relational arc memory.

Additionally, the system misclassified the "Roommates Dynamics" arc, which involves Meredith, Izzie, and George becoming roommates in the second episode. This storyline was grouped under a broader arc, "Intern Dynamics: Friendship And Rivalry." Human analysis, with its richer contextual understanding, deemed them distinct. This points to potential issues in the granularity of episodic memory encoding and the semantic interpretation by the LLM, which might over-generalize based on shared characters or settings.

Progressions within arcs were generally consistent with their titles and descriptions. However, they were not always fully captured for specific episodes, as episode summaries often omit minor developments that a human viewer, with a richer multimodal sensory input, might identify and store in their own memory. This highlights the dependency of the system's memory on the detail level of the input paratexts.

\subsection{Impact of the Graphic Interface on Memory Refinement}

The graphical interface proved highly effective in allowing human users to refine the system's narrative memory. It enabled human analysts to review, correct, and enhance extracted arcs efficiently. The LLM-assisted tools within the interface streamlined the correction process, allowing users to guide the "re-learning" or "re-consolidation" of narrative memories. This collaborative approach, where human intelligence curates and refines the computationally generated memory, underscores the value of human-computer interaction in complex narrative analysis \cite{piper2019enumerations}.

\section{\uppercase{Discussion: Memory Architectures and Computational Narratology}}
\label{sec:discussion_memory}

The development of our multi-agent system, grounded in principles of computational memory, offers several insights and raises important points for discussion within computational narratology and AI.

\subsection{Theoretical Contributions: MAS as a Narrative Memory Model}
Our system demonstrates a practical approach to implementing concepts from human memory theory—episodic, semantic, and working memory—within a computational framework for narrative analysis \cite{tulving1998episodic,baddeley2003working}. By architecting the MAS to explicitly manage different types of narrative information analogous to these memory systems, we provide a model for how AI can tackle the complexity of serialized stories.
\begin{itemize}
    \item The vector database as an \textit{episodic memory} store for specific narrative events allows for flexible, semantic retrieval of past occurrences, crucial for tracking arc development \cite{lewis2020retrieval}.
    \item The LLM as a \textit{semantic memory} resource provides the general knowledge about narrative structures, character roles, and thematic patterns necessary for interpreting and organizing these events \cite{radford2018improving}.
    \item The multi-agent workflow simulates \textit{working memory} operations, actively processing current inputs in conjunction with retrieved episodic data and semantic knowledge to build a coherent narrative understanding.
\end{itemize}
This approach moves beyond simple pattern matching, aiming for a more holistic, memory-based comprehension of narrative dynamics. It suggests that computational narratology can benefit from adopting such memory-centric architectures to better model the cumulative and interconnected nature of storytelling in series.

\subsection{Computational Memory vs. Human Narrative Memory}
While our system draws inspiration from human memory, it is crucial to acknowledge both similarities and fundamental differences.
\begin{itemize}
    \item \textbf{Advantages of Computational Memory}: The system possesses perfect recall of its encoded information (within the limits of its storage and retrieval algorithms), can process and compare numerous narrative elements simultaneously, and is not subject to the biases or decay that affect human memory \cite{schacter1999seven}. Its ability to systematically analyze every episode summary without fatigue is a clear advantage.
    \item \textbf{Limitations of Computational Memory}: Current computational systems, including ours, lack the rich and emotionally resonant memory of humans. Human understanding of narrative is deeply intertwined with personal experiences, cultural context, emotional responses, and Theory of Mind, aspects that are only superficially approached by LLMs \cite{herman2017storytelling}. Our system's reliance on textual paratexts means it misses subtext, visual cues, and performative nuances that heavily influence human interpretation and memory of a TV series.
\end{itemize}
The "memory" of our system is thus a functional analogue, designed for specific analytical tasks, rather than a replication of human cognitive processes.

\subsection{Implications for Narrative Analysis Tools}
The memory-grounded approach has several implications for the future development of narrative analysis tools:
\begin{itemize}
    \item \textbf{Enhanced Contextual Awareness}: Systems designed with explicit episodic and semantic memory components can maintain better contextual awareness over long narrative spans, leading to more coherent analyses.
    \item \textbf{Improved Handling of Seriality}: The ability to store, retrieve, and integrate narrative information across episodes is fundamental to understanding seriality.
    \item \textbf{Facilitating Human-AI Collaboration}: Interfaces that allow users to inspect, understand, and refine the system's "memory" (as demonstrated by our graphical interface) can lead to more powerful hybrid intelligence systems, combining computational scale with human insight \cite{piper2019enumerations}.
    \item \textbf{New Research Questions}: This framework opens up new research questions, such as how to model narrative forgetting or salience computationally, or how to better integrate different "sensory" (multimodal) inputs into a unified narrative memory.
\end{itemize}
Designing future tools with these memory principles in mind could lead to more sophisticated and insightful computational narratology.

\section{\uppercase{Conclusions and Future Work}}
\label{sec:conclusion}

This project has explored the application of a multi-agent system, conceptualized through the lens of computational memory architectures, to the complex task of extracting and analyzing narrative arcs in serialized television. By treating episode summaries as inputs to a system that extracts and processes episodic events via a multi-agent working memory, we have taken a step toward automating and structuring the analysis of intricate, temporally extended narratives.

The testing on \textit{Grey's Anatomy} demonstrated promising results, particularly in identifying discrete Anthology Arcs and character entities, where the demands on long-term memory integration are less severe. However, the reliance on textual paratexts exposed limitations in capturing the full complexity of overlapping storylines and subtle narrative dynamics that a human viewer, with richer multimodal input and more sophisticated inferential memory, would perceive. The system's current memory architecture, while functional, struggles with the nuanced consolidation and abstraction that characterizes human narrative recall, especially for interwoven, long-running arcs.

Despite these limitations, the memory-centric approach holds significant potential. The system is particularly well-suited for serialized written narratives—such as novels, episodic web fiction, or comic series—where the entirety of the narrative information resides within the text. In such formats, the "multimodal gap" is absent, allowing the system's text-focused memory components to operate more effectively.

Future work will aim to enhance the system's narrative memory capabilities. A key direction is the integration of multimodal inputs, such as subtitles (dialogue memory), scene descriptions (visual-spatial memory), and perhaps even analyses of visual or auditory data, to create a richer, more human-like episodic memory store. This would involve developing techniques for cross-modal memory linking and retrieval.

Refinements to the multi-agent system's working memory and memory consolidation processes are also needed. This includes improving the agents' ability to manage overlapping arcs, perform more robust deduplication, and better infer connections between seemingly disparate episodic events to form coherent long-term arc memories. Exploring more sophisticated memory update and forgetting mechanisms could also enhance realism and efficiency.

Expanding testing across a wider range of genres and narrative complexities will be crucial for refining the system's versatility and the generalizability of its memory architecture. Furthermore, investigating temporal memory dynamics more deeply—how the "memory" of an arc changes or fades in importance over time within the system—could yield insights into narrative salience. Exploring possibilities for cross-series memory transfer, where narrative patterns learned from one series inform the analysis of another, could be a long-term goal.

Ultimately, the combination of automated memory processing and human interpretive oversight remains indispensable. While computational systems excel at diligently storing and retrieving large volumes of narrative data, human intelligence continues to provide the crucial contextual understanding, inferential leaps, and critical judgment essential for meaningful narrative analysis. Future systems should aim to be even more transparent and interactive, functioning as true "cognitive partners" in the exploration of narrative memory.

\bibliographystyle{splncs04}
\bibliography{PAPER}

\end{document}